# Absence of an appreciable iron isotope effect on the transition temperature of the optimally doped SmFeAsO$_{1-y}$ superconductor

Parasharam M. Shirage[1,†], Kiichi Miyazawa[1,2], Kunihiro Kihou[1], Hijiri Kito[1,3], Yoshiyuki Yoshida[1], Yasumoto Tanaka[1], Hiroshi Eisaki[1,3], Akira Iyo[1,2,3,*]

[1] National Institute of Advanced Industrial Science and Technology, Tsukuba, Ibaraki 305-8568, Japan

[2] Department of Applied Electronics, Tokyo University of Science, 2641 Yamazaki, Noda, Chiba 278-3510, Japan

[3] JST, Transformative Research-Project on Iron Pnictides (TRIP), 5, Sanbancho, Chiyoda, Tokyo 102-0075, Japan

†E-mail: paras-shirage@aist.go.jp
*E-mail: iyo-akira@aist.go.jp

**Abstract:**

We report the iron (Fe) isotope effect on the transition temperature ($T_c$) in the oxygen-deficient $SmFeAsO_{1-y}$, a 50 K-class Fe-based superconductor. For the optimally-doped samples with $T_c$ = 54 K, change of the Fe average atomic mass ($M_{Fe}$) causes a negligibly small shift in $T_c$, with the Fe isotope coefficient ($\alpha_{Fe}$) as small as -0.024 ± 0.015 (where $\alpha_{Fe} = -d\ln T_c / d\ln M_{Fe}$). This result contrasts with the finite, inverse isotope shift observed in optimally-doped $(Ba,K)Fe_2As_2$, indicating that the contribution of the electron-phonon interaction markedly differs between these two Fe-based high-$T_c$ superconductors.

High-$T_c$ exceeding 50 K realized in the Fe-based superconductors has generated the worldwide extensive investigations towards discovering new materials with related structures, as well as towards elucidating their superconducting mechanism. With respect to the latter, the immediate question to be addressed is whether one can understand the high-$T_c$ superconductivity within the conventional framework, namely, the pairing of electrons mediated by the electron-phonon interaction, or more exotic mechanism should be introduced to account for the high-$T_c$ superconductivity.

It is well known that the isotope effect on $T_c$ has played a crucial role in establishing the basic concept of phonon-mediated superconductivity. In the framework of Bardeen-Cooper-Schrieffer's (BCS) theory, the isotope effect coefficient $\alpha$, defined by the relation $\alpha = -d\ln T_c / d\ln M$, is equal to 0.5 (in the case of compound superconductors, the sum of all isotope coefficients for the individual elements, $\alpha = \Sigma \alpha_i$, where $\alpha_i = -d\ln T_c / d\ln M_i$, becomes 0.5), which is indeed observed in many 'conventional' superconductors, such as elemental or binary alloy superconductors [1]. One example of such cases is MgB$_2$, a 38 K high-$T_c$ superconductor, in which boron (B) isotope exchange changes its $T_c$ by 1 K, corresponding to $\alpha_B$ = 0.26(3). The result clearly demonstrates that the superconductivity in MgB$_2$ is indeed mediated by the electron-phonon interaction [2]. On the other hand, in the optimally-doped cuprate superconductors, the lack of the oxygen isotope effect [3] has led the researchers to search for non-phonon-mediated mechanisms as a source of high-$T_c$ superconductivity. (It should be noted that the lack of the isotope effect itself does not necessarily imply the non-phonon mediated superconductivity. Indeed, several transition elements superconductors like Ru and Os show zero isotope coefficients, which are explained by strong electron-phonon coupling [4].)

In the Fe-based superconductors, their essential structural unit is the Fe-As plane. Accordingly, isotope experiments on the Fe or As site should provide crucial information to sort out the role of electron-phonon interaction in these systems. Based on this motivation, Fe isotope experiments were carried out by two groups. (Since As has only one stable isotope, As isotope experiment is practically impossible.) Both results indicated the finite isotope shift in $T_c$, although they contradict with each other. Liu *et al.* carried out $^{54}$Fe isotope exchange on SmFeAs(O,F) ($T_c \sim$ 42 K) and (Ba,K)Fe$_2$As$_2$ ($T_c \sim$ 38 K), and reported large isotope coefficients $\alpha_{Fe}$ = 0.34 and 0.37, respectively, which are comparable with the values expected from the BCS theory [5]. On the other hand, Shirage *et al.* carried out $^{54}$Fe and $^{57}$Fe isotope exchange on optimally-doped (Ba,K)Fe$_2$As$_2$ ($T_c \sim$ 38 K) and reported that heavier Fe isotope substitution results in increasing $T_c$, yielding the negative $\alpha_{Fe}$ = -0.18(3), referred to as "inverse" isotope effect [6]. The reason for the discrepancy is yet unclear. It may suggest that the isotope effect significantly depends on the subtle difference among the samples, such as carrier concentration, synthesis technique, lattice strain, *etc*.

In any case, results on the 40 K-class Fe-based superconductors indicate finite $\alpha_{Fe}$, strongly suggesting that the electron-phonon interaction exists and plays either positive [5] or negative [6] role. The natural question to be addressed is, whether the Fe isotope shift on $T_c$ also occurs in other Fe-based superconductors, in particular those with higher $T_c$'s. It is worth recalling that in the case of cuprates, the oxygen isotope effect is negligibly small in the optimally-doped samples, while the sizable isotope shift is observed for the under- or over-doped samples, as well as for the lower-$T_c$ materials [7-9]. There is also a possibility that $\alpha_{Fe}$ of the Fe-based superconductors exhibit significant material- and doping-dependence. Along the line of the reasoning, in this study we have examined the Fe isotope effect on a 50 K-class SmFeAsO$_{1-y}$ superconductor.

In order to evaluate the Fe isotope effect, one has to prepare the twin samples with different Fe isotopic mass under exactly the same conditions and then precisely compare their $T_c$'s. It is technically challenging since the shift in $T_c$ is estimated to be as small as 1.4 K, even if we assume the full isotope effect ($\alpha_{Fe} \sim 0.5$) when we substitute heavier $^{57}$Fe by lighter $^{54}$Fe. Moreover, $T_c$'s of the $Ln$FeAsO-based system strongly depend on their carrier concentration, which is changed by the amount of fluorine substitutions ($x$) in the case of $Ln$FeAsO$_{1-x}$F$_x$, or by the amount of oxygen deficiencies ($y$) for $Ln$FeAsO$_{1-y}$. It is critically important that the carrier concentrations of the twin samples are exactly the same and optimally-doped. In this regard, before carrying out the isotope exchange experiments, it is necessary to establish the relationship between $y$ and $T_c$, and to find out the optimal carrier concentrations.

In this study, polycrystalline samples of oxygen-deficient SmFeAsO$_{1-y}$ were prepared by high-pressure synthesis method as employed in our previous studies [10-12]. The starting materials were Fe, $^{54}$Fe$_2$O$_3$, $^{57}$Fe$_2$O$_3$, As, Sm(OH)$_3$ and a precursor of SmAs. (The role of Sm(OH)$_3$ addition is described below.) For the isotope substitution, we employed $^{57}$Fe$_2$O$_3$ and $^{54}$Fe$_2$O$_3$ powders, which were obtained by fully oxidizing the $^{57}$Fe enriched powder (ISOFLEX USA, enrichment 95.36%, average atomic mass 56.94, Fe purity 99.99%) and $^{54}$Fe enriched powder (ISOFLEX USA, enrichment 99.86%, average atomic mass 53.94, Fe purity 99.99%), respectively.

As a source of Fe, we used $^n$Fe (natural abundance, average atomic mass = 55.85) powder instead of using the raw (as-purchased) $^{57}$Fe and $^{54}$Fe metal powders, because the powders were partially oxidized from the beginning and the degree of oxidization were different from powder to powder. Even small oxidation of the starting elements increases the oxygen content $y$ in the synthesized materials, resulting in the shift of $T_c$ not due to the isotope exchange but due to the change in carrier concentration.

The materials of $^n$Fe, As, Sm(OH)$_3$ and SmAs were mixed and ground well using an agate mortar, and then divided into two equal parts. Then, the powder was mixed with the $^{57}$Fe$_2$O$_3$ (or $^{54}$Fe$_2$O$_3$) powders and ground again, then pressed into a pellet. These processes were carried out inside a glove box filled with dried N$_2$ gas. The two pellets were placed in a BN crucible and heated at about 1100°C under a pressure of about 3.5 GPa for 2 h. The sample assembly used for the sample synthesis is shown in the inset of Fig. 2. Powder X-ray diffraction (XRD) patterns of the samples were measured using CuK$_a$ radiation. The dc magnetic susceptibility was measured using a SQUID magnetometer (Quantum Design MPMS) under a nominal magnetic field of 5 Oe. The resistivity of the twin samples was measured simultaneously by a four-probe method using Quantum Design PPMS.

Introduction of the oxygen deficiency $y$ in $Ln$FeAsO$_{1-y}$ is reflected as the contraction of the $a$- and $c$- axis lattice parameters [13]. In Ref. [14], we reported that the addition of rare-earth hydroxide to the starting composition causes further contraction of the lattice parameters. Fig. 1 represents the $a$-parameter dependence of $T_c$ in SmFeAsO$_{1-y}$ and hydroxide added SmFeAsO$_{1-y}$(OH)$_x$. It shows that $T_c$ changes quite sensitively with the $a$-parameter. For the most underdoped samples with long $a$-axis length of 3.928 Å, $T_c$ is as low as 40 K. By introducing oxygen deficiency $y$, $T_c$ increases monotonously up to 54 K where the *a*-parameter becomes 3.915Å, the minimum length that we can reach. Then, by introducing Sm(OH)$_3$, additional decrease in the *a*-parameter is realized. Resultantly, a broad, flat-topped peak in $T_c$ (~ 54 K) shows up at around *a* = 3.913 Å. Existence of the peak in $T_c$ implies that there exists the 'optimal' carrier concentration in this system, as in other Fe-based superconductors such as (Ba,K)Fe$_2$As$_2$. The existence of this flat-top peak in $T_c$ allows us to extract the subtle change in $T_c$ caused by the isotope exchange, since the change in $T_c$ due to the fluctuation of carrier concentration is minimal around this broad peak.

Having established the ideal synthesis conditions for the isotope experiments, we synthesized the two samples with different Fe isotopes under exactly the same conditions. The nominal composition of the sample is SmFeAsO$_{0.77}$H$_{0.12}$, and 43 % of Fe in the samples is exchanged by the $^{57}$Fe or $^{54}$Fe isotopes. The resultant average atomic masses of Fe ($M_{Fe}$) are 56.32 and 55.02 for the samples synthesized using the $^{57}$Fe$_2$O$_3$ and $^{54}$Fe$_2$O$_3$ powders, respectively. If one assumes the isotope coefficient $\alpha_{Fe}$ to be 0.5, the shift in $T_c$ is 0.60 K, which is large enough to be detected experimentally.

XRD patterns show that the synthesized samples are single-phases within the experimental uncertainty (impurity phases less than about 5 % can not be detected) and the lattice parameters of the two samples synthesized at the same time are identical to each other. Fig. 2 shows the XRD patterns of the representative sample set (*S*1) as a typical example. The lattice parameters for the samples with the $M_{Fe}$ of 56.32 and 55.02 are *a* = 3.9110(2) Å, *c* = 8.4428(8) Å and *a* = 3.9111(2) Å, *c* = 8.4440(8) Å, respectively. The *a*-parameters of both samples correspond to the lattice spacing at which $T_c$ shows the maximum value in Fig. 1.

Fig. 3 (a) and (b) represents the temperature ($T$) dependence of the field-cooled (FC) magnetic susceptibility ($\chi$) for the two sample sets, $S1$ and $S2$, respectively. The magnitudes of $\chi$ at 5 K are around -0.01emu/g, corresponding to the superconducting volume fraction of 100 % without demagnetizing field correction. In Fig. 3, the magnitude of $\chi$ is normalized at their 5 K (-1) and 56 K (0) values for the sake of exact comparison. Sharp superconducting transition allows us to define the accurate $T_c$ value, as indicated in the inset of Fig. 3. Shifts in $T_c$ for the sample sets $S1$ and $S2$ are at most -0.05 K and -0.01 K, which correspond to $\alpha_{Fe}$ = -0.04 and -0.01, respectively.

Fig. 4 shows $T$ dependent resistivity of the sample set $S2$. Here again, the magnitude of the resistivity is normalized using the 300 K value. The resistivity data are almost overlapped with each other, indicative of the same carrier concentrations of the two samples. $T_c$'s of the samples are determined by the onset of resistive transition, which are 54.15 K and 54.10 K for $M_{Fe}$ = 56.32 and 55.02, respectively, as shown in the inset of Fig. 4. The estimated $\alpha_{Fe}$ is -0.04, which is comparable with that obtained from $\chi$.

Table I summarizes the $M_{Fe}$ dependence of $T_c$ determined from $\chi$ for five samples sets $S1$ - $S5$. The absolute value of $T_c$ is highly reproducible (54.02 K ± 0.13 K) and the difference in $T_c$ caused by the isotope substitution is at most -0.05 K, yielding the average $\alpha_{Fe}$ to be -0.024 ± 0.015. As mentioned, if we assume $\alpha_{Fe}$ as 0.5, the isotope shift is expected to be 0.6 K, which is one order of magnitude larger than the value obtained from the experiments. The results presented here evidently demonstrate that $\alpha_{Fe}$ is very small in the optimally doped, highest $T_c$ SmFeAsO$_{1-y}$ samples. Naively thinking, the negligibly small $\alpha_{Fe}$ may indicate that the superconductivity in this material is not caused by the electron-phonon interaction.

(At this moment there still exists the possibility that there is a finite isotope effect at As site.)

Our result is apparently in contradictory with a large $\alpha_{Fe}$ = 0.34 on SmFeAsO$_{0.85}$F$_{0.15}$ ($T_c$ = 41 K) reported by Liu *et al.* [5]. Giving lower $T_c$'s of their samples in comparison to the present ones, their samples are probably located at the under-doped region and the apparent difference might arise from the different carrier concentration of the samples. If this is the case, this situation resembles the cuprate superconductors in which the oxygen isotope effect is negligibly small in the optimal doped samples, while that in the under- or over-doped one is large, almost close to 0.5 [7-9].

Here we attempt to analyze the $\alpha_{Fe}$ of SmFeAsO$_{1-y}$ and (Ba,K)Fe$_2$As$_2$ by taking account of the possible competition between the contribution from phonons and antiferromagnetic (AF) spin fluctuations, as we have presented in the previous paper to explain the inverse isotope effect in (Ba,K)Fe$_2$As$_2$ [6,15]. According to this model, $T_c$ and $\alpha$ are described as:

$$k_B T_c = 1.13 \omega_{ph} \exp\left(-\frac{1}{\lambda_{ph} + \lambda^*_{AF}}\right) \quad \text{----(1)}$$

where $\lambda^*_{AF} = \dfrac{\lambda_{AF}}{1 - \lambda_{AF} \ln\left(\omega_{AF}/\omega_{ph}\right)}$ ----(2)

$$\alpha = \frac{1}{2}\left(1 - \left(\frac{\lambda^*_{AF}}{\lambda_{ph} + \lambda^*_{AF}}\right)^2\right) \quad \text{----(3)}$$

where $\lambda_{AF}$, $\omega_{AF}$, $\lambda_{ph}$, $\omega_{ph}$ are the dimensionless coupling constants and characteristic energies of the AF spin fluctuation and phonon, respectively. $\lambda_{ph}$ is defined by $\lambda_{ph}^{intraband}$ - $\lambda_{ph}^{interband}$, where $\lambda_{ph}^{intraband}$ and $\lambda_{ph}^{interband}$ are the dimensionless electron-phonon coupling constants for intraband and interband pair hoppings, respectively. In the present calculation we provided [16] the value of $\omega_{ph}$ = 350 K. The $\lambda^*_{AF}$ and $\lambda_{ph}$ deduced from eqs. (1) and (3) by using the observed $T_c$ and $\alpha_{Fe}$ values are listed in Table II.

One can see that the values of $\lambda^*_{AF}$ are almost the same for both systems at their optimal $T_c$. Within the present framework, the higher $T_c$ for SmFeAsO$_{1-y}$ in comparison to (Ba,K)Fe$_2$As$_2$ is ascribed to the smaller magnitude of $\lambda_{ph}$, namely, the larger (negative) contribution of the intraband electron-phonon coupling, rather than the larger $\lambda^*_{AF}$. The lack of the isotope effect in the optimal doped SmFeAsO$_{1-y}$ implies that the $\lambda_{ph}^{interband}$ is comparable to the $\lambda_{ph}^{intraband}$. According to this model, $T_c$ can be enhanced if the intraband electron-phonon interaction works cooperatively with the AF spin fluctuation ($\lambda_{ph}^{intraband} > \lambda_{ph}^{interband}$).

In summary, we have studied the Fe isotope effect on $T_c$ in the optimally-doped SmFeAsO$_{1-y}$ ($T_c$ = 54 K). Change of the $M_{Fe}$ yields a negligibly small shift in $T_c$, with its $\alpha_{Fe}$ as small as -0.024 ± 0.015. This result indicates that $\alpha_{Fe}$ strongly depends on the system and carrier concentrations. To clarify the issue, it is desirable to evaluate $\alpha_{Fe}$ as a function of doping level for Fe-based systems in the future.

This work was supported by a Grant-in-Aid for Specially Promoted Research (20001004) from the Ministry of Education, Culture, Sports, Science and Technology (MEXT), and Mitsubishi Foundation.

Table I. The Fe average atomic mass ($M_{Fe}$) dependence of the $T_c$ determined from $\chi$ for five sample sets $S1$ - $S5$.

| Sample sets | $M_{Fe}$ | $T_{c(\chi)}$ (K) | $\Delta T_{c(\chi)}$ (K) | $\alpha_{Fe}$ |
|---|---|---|---|---|
| $S1$ | 55.02 | 54.02 | -0.05 | -0.040 |
|  | 56.32 | 54.07 |  |  |
| $S2$ | 55.02 | 53.79 | -0.01 | -0.008 |
|  | 56.32 | 53.80 |  |  |
| $S3$ | 55.02 | 53.96 | -0.02 | -0.016 |
|  | 56.32 | 53.98 |  |  |
| $S4$ | 55.05 | 54.10 | -0.02 | -0.016 |
|  | 56.31 | 54.12 |  |  |
| $S5$ | 55.02 | 54.13 | -0.05 | -0.040 |
|  | 56.32 | 54.18 |  |  |
|  |  |  | *Avarage* $\alpha_{Fe}$ | -0.024 ± 0.015 |

Table II. Deduced $\lambda_{AF}$ and $\lambda_{ph}$ from eqs. (1) and (3) using the observed $T_c$ and $\alpha_{Fe}$ values for (Ba,K)Fe$_2$As$_2$ and SmFeAsO$_{1-y}$.

| | Observed | | Provided | Deduced | | |
|---|---|---|---|---|---|---|
| | $T_c$ (K) | $\alpha_{Fe}$ | $\omega_{ph}$ (K) | $\lambda_{ph}/\lambda^*_{AF}$ | $\lambda^*_{AF}$ | $\lambda_{ph}$ |
| (Ba,K)Fe$_2$As$_2$ | 38 | $-0.18 \pm 0.03$ | 350 | $-0.14 \pm 0.02$ | $0.50 \pm 0.01$ | $-0.07 \pm 0.01$ |
| SmFeAsO$_{1-y}$ | 54 | $-0.02 \pm 0.01$ | 350 | $-0.02 \pm 0.01$ | $0.51 \pm 0.01$ | $-0.01 \pm 0.005$ |

**Figure captions**

Figure 1. Dependence of $T_c$ on the lattice parameter $a$ in oxygen-deficient SmFeAsO$_{1-y}$ and hydroxide added SmFeAsO$_{1-y}$(OH)$_x$.

Figure 2. X-ray diffraction patterns of the sample set $S$1 with $M_{Fe}$ of 55.02 and 56.32 synthesized simultaneously under high pressure. The inset shows the sample cell assembly for the high-pressure synthesis method.

Figure 3. Temperature ($T$) dependence of the field-cooled (FC) susceptibility ($\chi$) for (a) sample set $S$1 and (b) sample set $S$2 normalized at 5 K and 56 K. Inset shows the $T$ dependence of $\chi$ near $T_c$. The Fe isotope coefficient is estimated to be $\alpha_{Fe}$ = -0.04 and -0.01 for the $S$1 and $S$2, respectively.

Figure 4. Temperature ($T$) dependence of resistivity ($\rho$) of the sample set $S$2 with $M_{Fe}$ of 55.02 and 56.32. The inset shows the $T$ dependence of $\rho$ near $T_c$. The Fe isotope coefficient is estimated to be $\alpha_{Fe}$ = -0.04.

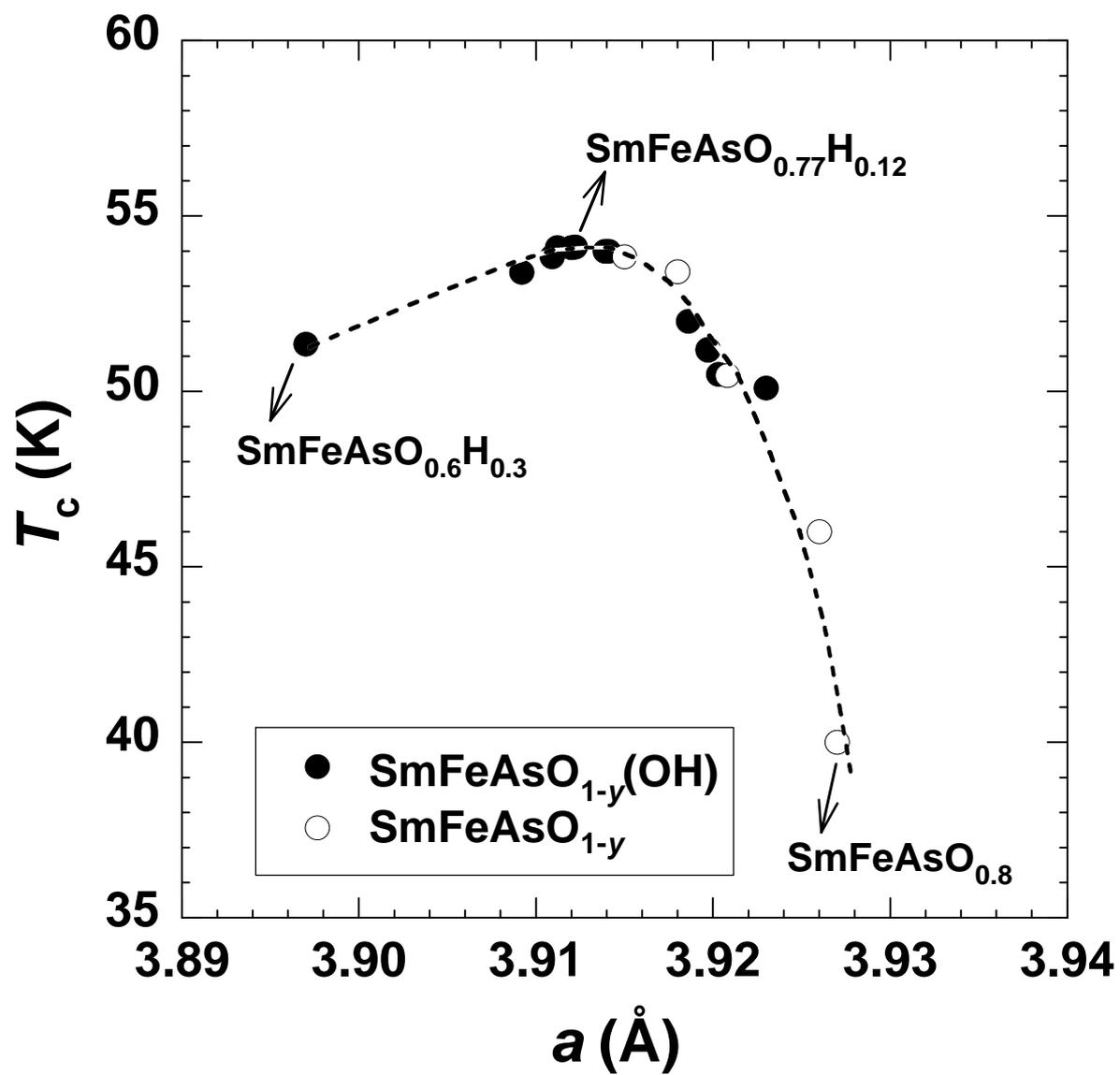

**Figure 1**

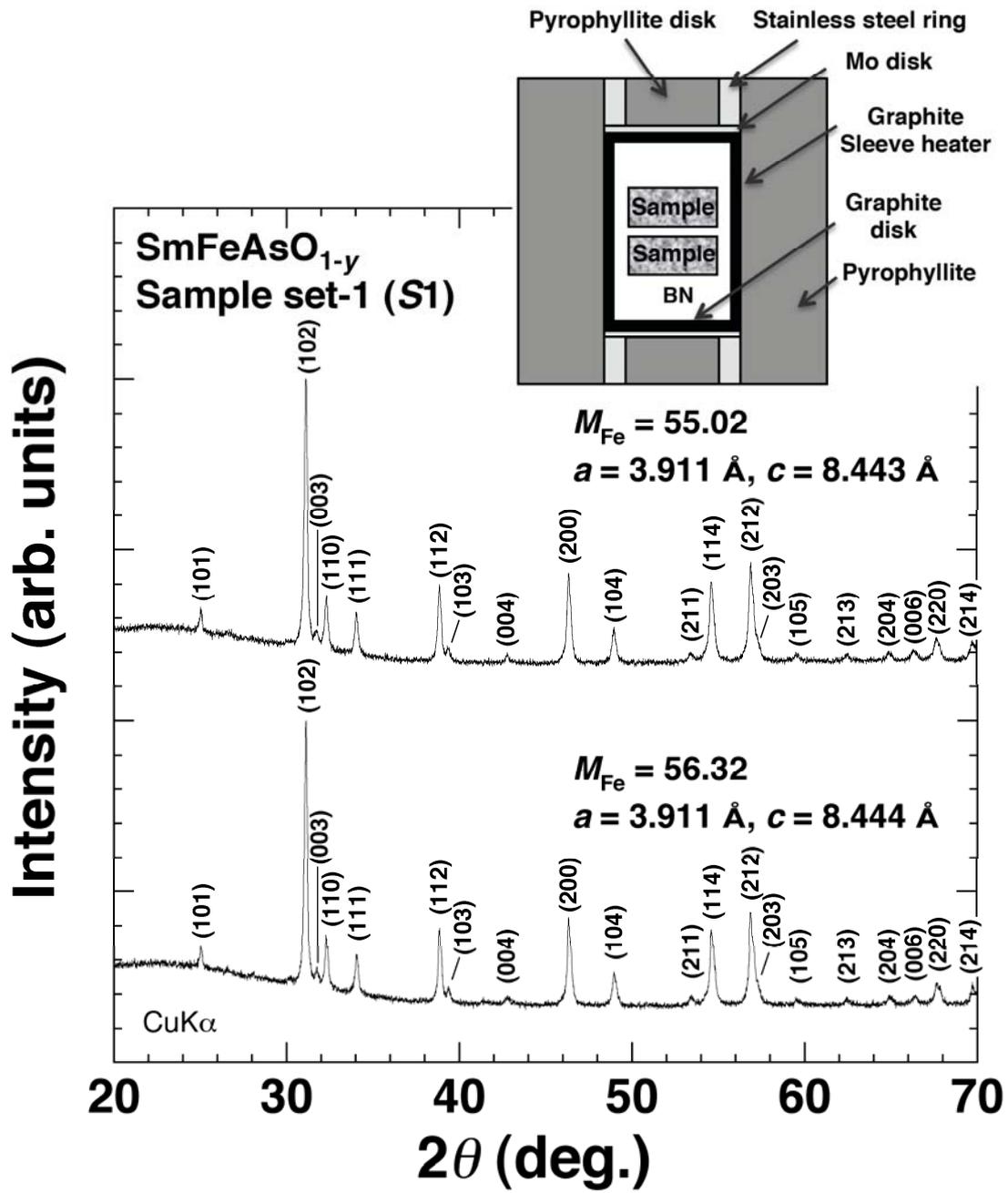

**Figure 2**

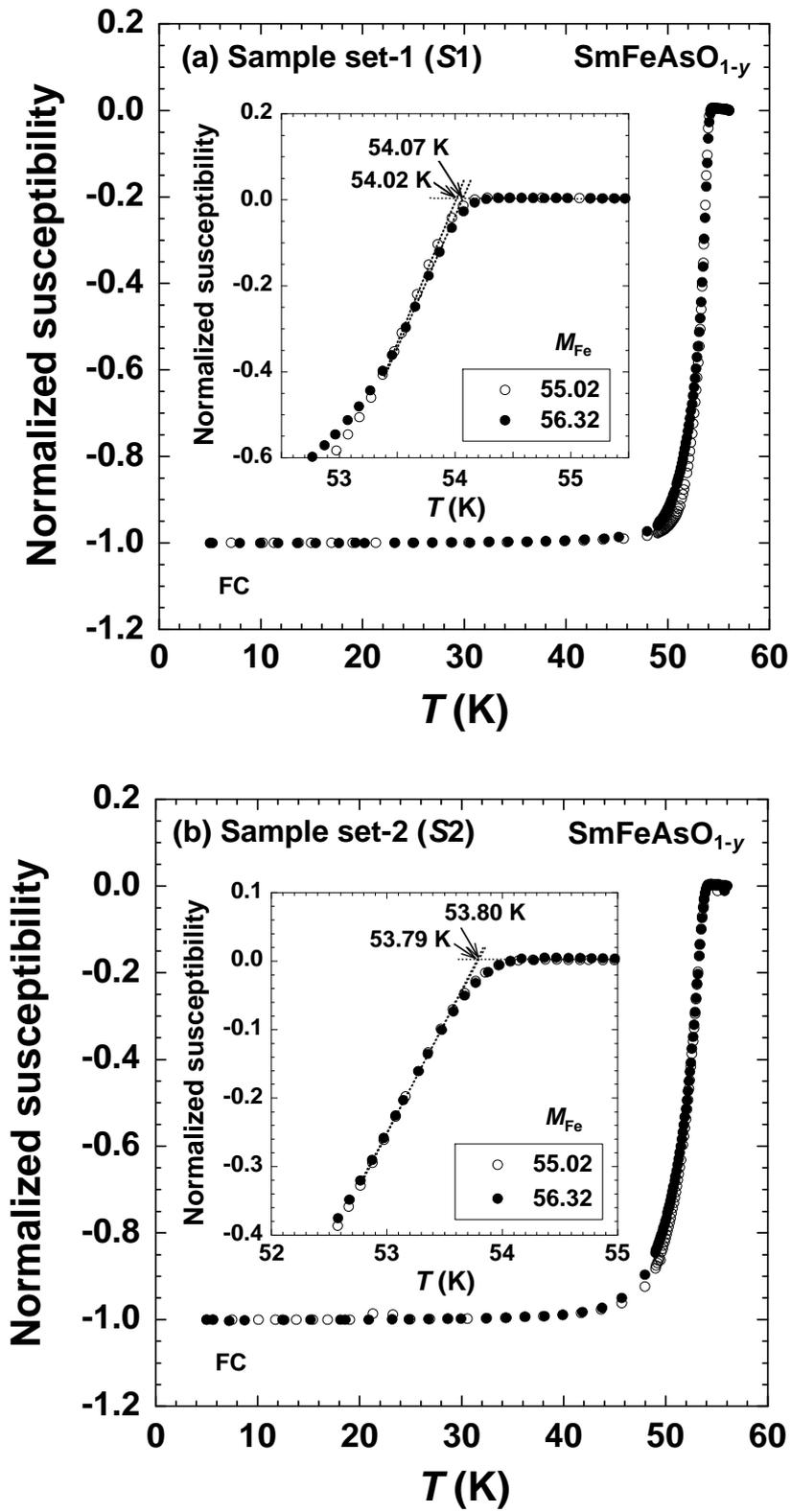

**Figure 3**

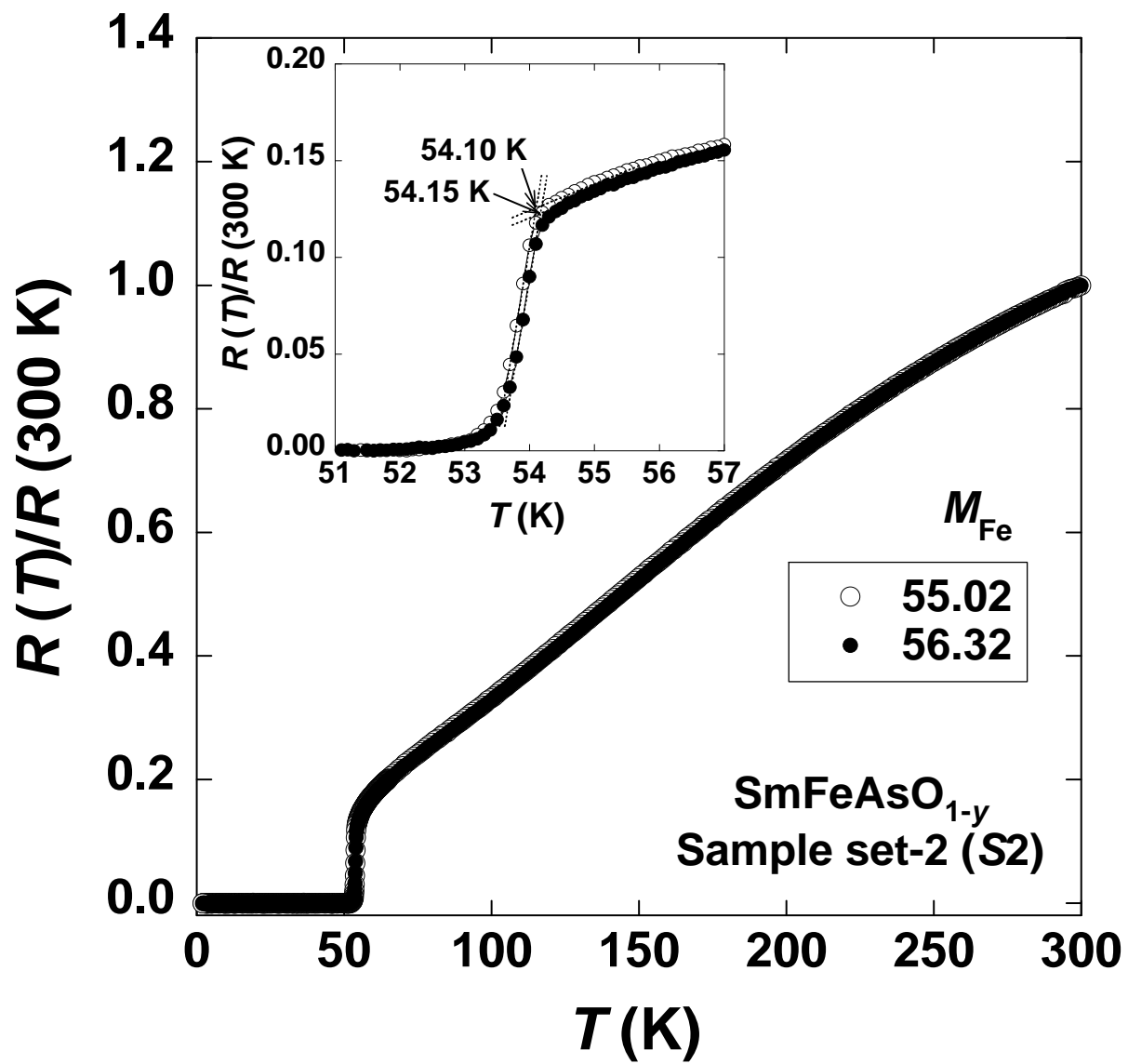

**Figure 4**